\documentclass[traditabstract]{aa}
\usepackage{graphicx}
\usepackage{natbib}
\usepackage{amssymb}
\usepackage{amsfonts}
\usepackage{amsbsy}
\usepackage{amsmath}

\bibpunct{(}{)}{;}{a}{}{,}

\newcommand{\Scmb}{\mathfrak{S}_{\rm c}(p)}

\newcommand{\Scbj}{\boldsymbol{\mathfrak{S}}_j}

\newcommand{\Scba}{\boldsymbol{\mathfrak{S}}_1}
\newcommand{\Scbb}{\boldsymbol{\mathfrak{S}}_2}
\newcommand{\ScbNc}{\boldsymbol{\mathfrak{S}}_{Nc}}
\newcommand{\Scmbh}{\widehat{\mathfrak{S}}_{\rm c}(p)}
\newcommand{\Scmbhb}{\widehat{\boldsymbol{\mathfrak{S}}}_{\rm c}}
\newcommand{\Scmbb}{\boldsymbol{\mathfrak{S}}_{\rm c}}

\newcommand{\Smthb}{\boldsymbol{\mathfrak S}}

\newcommand{\Sgalb}{\boldsymbol{S}_{\rm f}}
\newcommand{\Sgalib}{\boldsymbol{S}^{(i)}_{\rm f}}
\newcommand{\Sib}{\boldsymbol{S}^{(i)}}
\newcommand{\Sgal}{S^{(i)}_{\rm f}(p)}
\newcommand{\Si}{S^{(i)}(p)}

\newcommand{\Nmthi}{{\mathcal N}^{(i)}}

\newcommand{\Nmthb}{\boldsymbol{\mathcal N}}

\newcommand{\oneb}{\boldsymbol{1}}
\newcommand{\Ab}{\boldsymbol{A}}

\newcommand{\ab}{\boldsymbol{a}}
\newcommand{\bb}{\boldsymbol{b}}

\newcommand{\Cb}{\boldsymbol{C}}

\newcommand{\Sb}{\boldsymbol{S}}

\newcommand{\wb}{\boldsymbol{w}}
\newcommand{\zerob}{\boldsymbol{0}}

\begin{document}

   \title{Considerations on some neglected but impotant issues concerning the Internal Linear Combination method in Astronomy}
   \author{R. Vio\inst{1}
    \and
           P. Andreani\inst{2}
          }
   \institute{Chip Computers Consulting s.r.l., Viale Don L.~Sturzo 82,
              S.Liberale di Marcon, 30020 Venice, Italy\\
              \email{robertovio@tin.it},
         \and
		  ESO, Karl Schwarzschild strasse 2, 85748 Garching, Germany\\
                  INAF-Osservatorio Astronomico di Trieste, via Tiepolo 11, 34143 Trieste, Italy\\              
		  \email{pandrean@eso.org}
             }

\date{Received .............; accepted ................}

\abstract
{Although the {{\it internal linear combination}} method (ILC) is a technique widely used for the separation of the Cosmic Microwave
Background signal from the Galactic foregrounds, its characteristics are not yet well defined. This can lead to misleading conclusions
about the actual potentialities and limits of such approach in real applications. Here we discuss briefly some facts about ILC that to
our knowledge are not fully worked out in literature and yet have deep effects in the interpretation of the results.} 
\titlerunning{ILC method}
\authorrunning{R. Vio, \& P. Andreani}
\maketitle

\section{Introduction}

A widely used approach for the separation of the {\it cosmic microwave background} (CMB) from the diffuse Galactic background is
the {\it internal linear combination} method (ILC).
For instance, this method was adopted in the reduction of the data from the Wilkinson Microwave Anisotropy Probe (WMAP) satellite for CMB observations 
\citep{ben03}. Its success is due to the fact that, among the separation techniques, ILC calls for the smallest number of {\it a priori} 
assumptions.
If the data are in the form of $N_o$ maps, taken at different frequencies and containing $N_p$ pixels each, the model
on which ILC is based is
\begin{equation} \label{eq:image}
\Si = \Scmb + \Sgal + \Nmthi(p).
\end{equation}
Here, $\Si$ provides the value of the $p$th pixel for a map obtained at channel ``$~i~$'' \footnote{In the present work, $p$ 
indexes pixels in the classic spatial domain. However, the same formalism applies if other domains are considered, for example, the Fourier one.}, 
whereas $\Scmb$, $\Sgal$ and $\Nmthi(p)$ are the contributions due to the CMB, the diffuse Galactic foreground and the experimental noise, 
respectively. Although not necessary, often it is assumed that all of these contributions are representable by means of stationary 
random fields. Moreover, without loss of generality, for ease of notation the random fields are supposed as the
realization of zero-mean spatial processes. 
The basic idea behind model~(\ref{eq:image}) is that, contrary to the components that form the Galactic background, CMB is independent of the 
observing channel. 
ILC exploits this fact 
averaging $N_o$ images $\{ \Si \}_{i=1}^{N_o}$ and giving a specific weight $w_i$ to each of them so as to minimize the 
impact of the foreground and noise
\citep{ben03}. This means to look for a solution of type
\begin{equation} \label{eq:sum}
\Scmbh = \sum_{i=1}^{N_o} w_i \Si.
\end{equation}
If the constraint $\sum_{i=1}^{N_o} w_i = 1$ is imposed, Eq.~(\ref{eq:sum}) becomes
\begin{equation} \label{eq:wls}
\Scmbh = \Scmb + \sum_{i=1}^{N_o}  w_i [ \Sgal + \Nmthi(p) ]. 
\end{equation}
Now, from this equation it is clear that, for a given pixel ``$p$'',  the only variable terms are in the sum. Hence,  
under the assumption of independence of $\Scmb$ from $\Sgal$ and $\Nmthi(p)$,
the weights $\{ w_i \}$ have to minimize the variance of $\Scmbh$, i.e.
\begin{align}
& \{ w_i \} = \underset{\{ w_i \} }{\arg\min} \nonumber \\
& {\rm VAR} \left[\Scmb \right] + {\rm VAR}\left[\sum_{i=1}^{N_o}  w_i (\Sgal + \Nmthi(p)) \right],
\end{align}
where ${\rm VAR}[s(p)]$ is the {\it expected variance} of $s(p)$.
If $\Sib$ denotes a {\bf row vector} such as $\Sib = [S^{(i)}(1), S^{(i)}(2), \ldots, S^{(i)}(N_p)]$ and the $N_o \times N_p$ matrix $\Sb$ is defined as  
\begin{equation}
\Sb = 
\left( \begin{array}{c}
\Sb^{(1)} \\
\Sb^{(2)} \\
\vdots \\
\Sb^{(N_o)}
\end{array} \right),
\end{equation}
then Eq.~(\ref{eq:image}) becomes
\begin{equation} \label{eq:basicmm}
\Sb = \Scmbb + \Sgalb + \Nmthb.
\end{equation}
In this case, the weights are given by \citep{eri04}
\begin{equation} \label{eq:wr}
\wb = \frac{\Cb_{\Sb}^{-1} \oneb}{\oneb^T \Cb_{\Sb}^{-1} \oneb},
\end{equation}
where 
$\Cb_{\Sb}$ is the $N_o \times N_o$ cross-covariance matrix of the random processes that generate $\Sb$, i.e.
\begin{equation}
\Cb_{\Sb} = {\rm E}[\Sb \Sb^T], 
\end{equation}
and $\oneb = (1, 1, \ldots, 1)^T$ is a column vector of all ones. Here, ${\rm E}[.]$ denotes the {\it expectation operator}.  
Hence, the ILC estimator takes the form
\begin{align} 
\Scmbhb & = \wb^T \Sb, \label{eq:wlss} \\
        & = \alpha \oneb^T \Cb_{\Sb}^{-1} \Sb, \label{eq:basic}
\end{align}
with 
$\oneb^T \wb = 1$ and the scalar quantity $\alpha$ given by
\begin{equation} \label{eq:alpha}
\alpha = [ \oneb^T \Cb_{\Sb}^{-1} \oneb]^{-1}.
\end{equation}

In practical applications, matrix $\Cb_{\Sb}$ is unknown and has to be estimated from the data. Typically, this is done by means of the
estimator
\begin{equation} \label{eq:C}
\widehat \Cb_{\Sb} = \frac{1}{N_p} \Sb \Sb^T.
\end{equation}
In this case, the ILC estimator is given by Eqs.(\ref{eq:wlss})-(\ref{eq:alpha}) with $\Cb_{\Sb}$ and $\wb$ replaced, respectively, by 
$\widehat \Cb_{\Sb}$ and
\begin{equation} \label{eq:w}
\widehat \wb = \frac{\widehat \Cb_{\Sb}^{-1} \oneb}{\oneb^T \widehat \Cb_{\Sb}^{-1} \oneb}.
\end{equation}

\section{Some unfocussed points about ILC}

In spite of its popularity, various questions concerning ILC appear not yet well fixed. A first issue is linked to the fact
that the estimation of the power-spectrum of CMB is a problem deeply different from the separation of this component from the Galactic foreground. 
Indeed, the estimation of the second-order properties of a stochastic signal, though contaminated by noise, is an easier task than
its recovery. Here, the point is that, if one is interested in the spatial distribution of the CMB emission, then in Eq.~(\ref{eq:image}) 
the term $\Sgal + \Nmthi(p)$ cannot be considered as a single noise 
component. Often such assumption is made \citep[e.g., see ][]{eri04,hin07,del09,dic09} since
in this way the problem is reduced to the separation of two components only and no {\it a priori} information
on the ``global'' noise is required. Actually, this procedure can lead to wrong conclusions. 
For example, since all of the component in the mixtures $\Sb$ are supposed to be the realization of a zero-mean random processes, 
from Eq.~(\ref{eq:wls}) one could derive that
\begin{equation}
{\rm E}[\Scmbhb | \Scmbb] = \Scmbb + \wb^T {\rm E}[\Sgalb] + \wb^T {\rm E}[\Nmthb] = \Scmbb,
\end{equation}
i.e. the ILC estimator is unbiased \footnote{Here, ${\rm E}[\ab|\bb] = \left[ {\rm E}[a(1) | b(1)], {\rm E}[a(2) | b(2)], \ldots, 
{\rm E}[a(N_p) | b(N_p)] \right]^T$ 
with ${\rm E}[a(p) | b(p)]$ the {\it conditional expectation}
of $a(p)$ given $b(p)$.}. This is not correct: the claim that $\Scmbhb$ is unbiased requires to prove that
\begin{equation}
{\rm E}[\Scmbhb | \Scmbb, \Sgalb] = \Scmbb + \wb^T \Sgalb + \wb^T {\rm E}[\Nmthb] = \Scmbb.
\end{equation}
The reason is that $\Sgalb$ is a {\bf fixed} realization of a random process. There is no way to obtain another one. 
Even if observed many times (under the same experimental conditions) the foreground components (for instance the Galaxy)
will always appear the same.
Only the noise component $\Nmthb$ will change. Indeed, apart from very specific situations, in general ILC has to be expected to provide biased
estimates of CMB \citep{vio08}. Here, we stress that the biased results provided by ILC is a well established fact only relatively
to the estimation of the power-spectrum of CMB \citep[e.g.][]{sah08}.

A second issue, often neglected in literature, is the belief that, in order to work, ILC does not require any hypothesis 
about $\Sgalb$. 
Again, this is not correct. As proved by \citet{vio08}, even in the case of noise-free data, one may hope to obtain an unbiased estimate of the CMB only
if $\Sgalb$ is given by a linear mixture of the contribution of $N_c$ physical processes $\{ \Scbj \}_{j=1}^{N_c}$, i.e.
\begin{equation} \label{eq:gal}
\Sgalib = \sum_{j=1}^{N_c} a_{ij} \Scbj,
\end{equation}
with $a_{ij}$ constant coefficients and $N_c < N_o$. This means that for the $j$th physical process a template $\Scbj$ 
is assumed to exists which is independent of the specific channel ``$~i~$''. Moreover, the number of these channels has to be greater than that of
the physical processes. Inserting Eq.~(\ref{eq:gal}) into Eq.~(\ref{eq:basicmm}) one obtains
\begin{equation} \label{eq:model}
\Sb = \Ab \Smthb + \Nmthb
\end{equation}
with
\begin{equation} \label{eq:modelS}
\Smthb = \left( \begin{array}{l}
\Scmbb \\
\Scba \\
\Scbb \\
\vdots \\
\ScbNc \\
\end{array} \right),
\end{equation}
and
\begin{equation}
\Ab = 
\left( \begin{array}{cccccc} \label{eq:Amatrix}
1 & \vline & a_{11} & a_{12} & \ldots & a_{1 N_c} \\
\hline
1 & \vline & a_{21} & a_{22} & \ldots & a_{2 N_c} \\
\vdots & \vline & \vdots & \vdots & \ddots & \vdots \\
1 & \vline & a_{N_o 1} & a_{N_o 2} & \ldots & a_{N_o N_c}
\end{array} \right).
\end{equation}
This is the only model that ILC is able to handle \footnote{If data are contaminated by noise, i.e. $\Nmthb\neq \zerob$,
then in Eq.~(\ref{eq:wr}) $\Cb_{\Sb}$ has to be substituted with $\Cb_{\Sb} - \Cb_{\Nmthb}$.}. This fact should suggest extreme caution
when dealing with questions as the nonGaussianity of CMB. Indeed, the results produced by ILC are extremely sensitive to the presence of
components that are not modelizable by means of Eqs.~(\ref{eq:model})-(\ref{eq:Amatrix}). By instance, without renouncing the cosmic origin of CMB,
this could explain the spatial associations claimed by \citet{ver07} between the interstellar neutral hydrogen (HI) emission morphology and 
small-scale structure observed by the Wilkinson Microwave Anisotropy Probe (WMAP). 

\begin{figure*}
        \resizebox{\hsize}{!}{\includegraphics{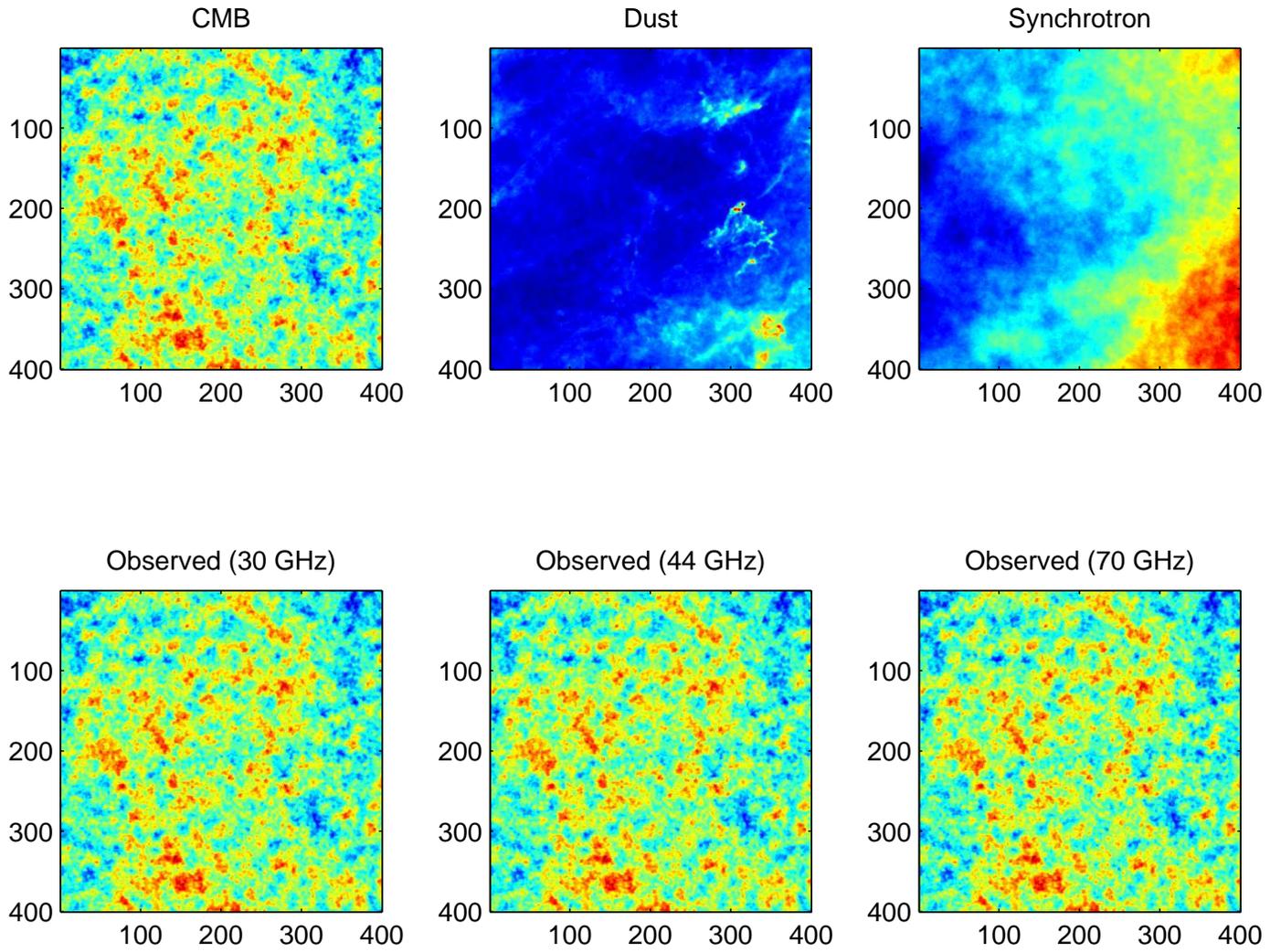}}
        \caption{Original templates (top panels) and the corresponding three mixtures (bottom panels) used in the numerical experiments described in the 
         text.}
        \label{fig:fig1}
\end{figure*}

\begin{figure*}
        \resizebox{\hsize}{!}{\includegraphics{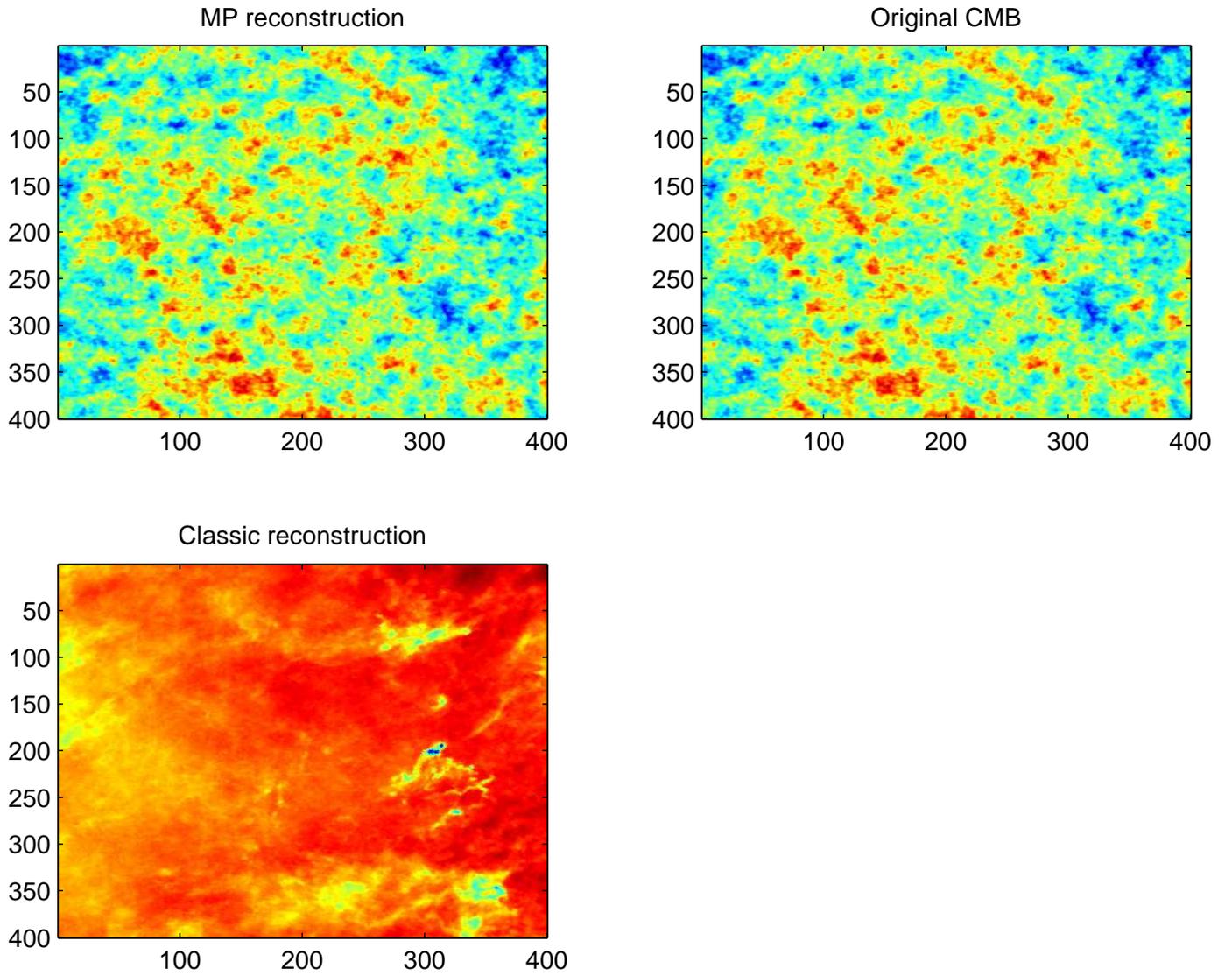}}
        \caption{ILC solutions obtained using, respectively, the {\it Moore-Penrose} inverse of $\widehat\Cb_{\Sb}$ (top-left panel) and
        the original weights~(\ref{eq:w}) (bottom-left panel). For reference, the target CMB component is also presented (top-right panel).}
        \label{fig:fig2}
\end{figure*}

A third issue is that, using the weights $\wb$, it is implicitly assumed that
$\Cb_{\Sb}$ is a well 
conditioned matrix. However, there are various practical situations where this condition is not satisfied. By instance, this can
be expected for high signal-to-noise ratio observations at high Galactic latitude, i.e. when noise $\Nmthb$ is negligible and
CMB is by far the dominant component with $a_{ij} \ll 1$, $i = 1, 2, \ldots, N_o$, $j=1, 2, \ldots, N_c$. Indeed, since
\begin{equation} \label{eq:cov}
\Cb_{\Sb} = \Ab \Cb_{\Smthb} \Ab^T,
\end{equation}
when $\Ab$ is badly conditioned it has to be expected that the same holds for $\Cb_{\Sb}$. 
As a consequence, an imprecise calibration of the $N_o$ maps, in such a way that the true value of the entries in the first 
column of $\Ab$ are different from ``$1$'',
will be amplified with catastrophic consequences. This is what lead \citet{dic09} to the conclusion that, in the case of high 
signal-to-noise ratio observations,
ILC is extremely sensitive to calibration errors. If true, this should be a quite troublesome situation since it limits the
usefulness of the observations in a high signal-to-noise ratio regime. Actually, this problem can be easily avoided if in Eq.~(\ref{eq:wr})
$\Cb^{-1}_{\Sb}$ is substituted with the corresponding {\it Moore-Penrose} inverse $\Cb^{\dag}_{\Sb}$. These arguments hold also when the weights
$\widehat\wb$ are used and $\widehat\Cb^{\dag}_{\Sb}$ substitutes $\widehat\Cb^{-1}_{\Sb}$ in Eq.~(\ref{eq:w}).
This is clearly visible in Figs.~\ref{fig:fig1} and \ref{fig:fig2} that show the results
provided by ILC in the case of a simulated high Galactic latitude observation
at three different frequencies, say $30$, $44$, and $70$ ${\rm GHz}$. The region analyzed
is a square patch ($400 \times 400$ pixels) with side
of about $24^{\circ}$, centered
at $l=90^{\circ}$, $b=45^{\circ}$ (Galactic coordinates). Three components have been considered, i.e. CMB, synchrotron and dust.
More details can be found in \citet{vio03}.
The effects of the different point spread function for the various frequencies have been neglected.
The top panels of Fig.~\ref{fig:fig1} shows the templates relative to the phsyical processes that have been used to form the three
linear mixtures shown in the bottom panel of the same figure. No noise has been added. Perfect frequency scaling has been assumed, i.e.
model~(\ref{eq:model})-(\ref{eq:Amatrix}) holds, but a calibration error of $10\%$ has been imposed in all of the channels. The assumption of
perfect frequency scaling constitutes only an approximation. However, since CMB 
by far constitutes the dominant component, this is of secondary 
importance. In fact, the condition number of the resulting $\widehat\Cb_{\Sb}$ is
$1.6 \times 10^6$, i.e. this matrix is ill-conditioned. The results obtainable by ILC using the original weights~(\ref{eq:w})
and their version computed using $\widehat\Cb^{\dag}_{\Sb}$ are shown in Fig.\ref{fig:fig2}. It is evident
that, if not properly addressed, the ill-conditioning of $\widehat\Cb_{\Sb}$ has catastrophic consequences on the quality of the separation.

\section{Conclusions} 
From the analysis presented above, it is evident that the results provided by ILC have to be interpreted with extreme caution. This techniques suffers
many drawbacks that, if not properly taken into account, can lead to misleading if not wrong conclusions. In particular, ILC should be used
only in situations where matrix $\widehat\Cb_{\Sb}$ is (close to be) singular, i.e. only when one can be certain
that model~(\ref{eq:model})-(\ref{eq:Amatrix}) holds with $N_o > N_c+1$. In the contrary case, the separation operated through ILC has to 
be expected inaccurate.

\clearpage
\clearpage

\end{document}